\documentclass[aps,prb,showpacs]{revtex4}

\usepackage{amsmath}

\begin{document}

\input psfig.sty

\title{On magnetoconductivity of metallic manganite phases and heterostructures}
\author{M.\ Dzero
\footnote{Also at: Physics Department, Florida State University.
Electronic address: dzero@magnet.fsu.edu}$^{,1}$,
L.\ P.\ Gor'kov$^{1,2}$, and V.\ Z.\ Kresin$^{3}$}
\affiliation{$^1$National High Magnetic Field Laboratory, Florida State
University, Tallahassee, FL 32310}
\affiliation{$^2$L.D. Landau Institute for Theoretical Physics, Russian Academy
of Sciences, 117334 Moscow, Russia}
\affiliation
{$^3$Lawrence Berkeley Laboratory, University of California, Berkeley,
CA 94720}

\begin{abstract}
We use the double exchange (DE) model via degenerate orbitals and tight-binding
approximation to study the magnetoconductivity of a canted A-phase of
pseudo-cubic manganites. It is argued that the model is applicable in a broad
concentration range for manganites A$_{1-x}$B$_x$MnO$_3$ with the tolerance factor,
$t$, close to one. As for the substitutional disorder, scattering on random Jahn-Teller
distortions of MnO$_6$ octahedra is chosen. We emphasize an intimate correlation
between the carrier concentration and resistivity value of metallic manganites.
Magnetoresistance as a function of magnetization is calculated for a canted A-phase for
both in-plane and out-of-plane current directions. A contact between two manganite phases
is considered and structure of the transition region near the contact is discussed. Numerical
calculations show charge re-distribution near the contact and a large screening length
of the order of five inter-atomic distances. We employed our results to interpret data obtained
in recent experiments on La$_{0.4}$Sr$_{0.6}$MnO$_3$/La$_{0.55}$Sr$_{0.45}$MnO$_3$ superlattices.
We also briefly discuss the relative importance of the cooperative Jahn-Teller distortions, double
exchange mechanism and super-exchange interactions for the formation of the A-phase at increasing
Sr concentrations $x>0.45$ in  La$_{1-x}$Sr$_{x}$MnO$_3$ to suggest that the Jahn-Teller contraction
of octahedra, $c/a<1$, plays a prevailing role.
\end{abstract}


\maketitle

\section{Introduction}
Pseudo-cubic manganites are remarkable for their rich phase
diagram although properties of various phases have not been
studied equally well. For example, the mechanisms of the "charge
ordering" phenomenon (CO-phase) or of the metallic
anti-ferromagnetic phase (A-phase), which usually appear near
$x\simeq{0.5}$, to a large extent remain to be poorly understood.
It is also surprising that the potential of metallic manganites as the
basic elements for various hetero-structures in development of
alternate devices is practically unexplored. Meanwhile the
diversity of their phases depending on temperature and doping
concentration poses interesting questions regarding phenomena that
may take place at the interface of the contacts between them.
Further understanding of magneto-transport properties of these
materials also remains of prime importance because of several
reasons. One of them is the colossal
magneto-resistance effect, observed in these compounds at room
temperatures. At the same time there is an increased general
interest in the artificial engineering of nano-structure materials
with novel physical properties, such as the giant
magneto-resistance (GMR) effect (see \cite{Parkin} for a review).
The discovery of the GMR effect has generated a lot of
investigations focused on understanding the physical mechanism of
this phenomena and that is why use of artificial hetero-structures
composed of manganites suggests an interesting direction of
research. Indeed, the A-phase with its weakly coupled ferromagnetic planes
is a natural spin-valve system itself ~\cite{Nd}. Recent experimental 
results for large magneto-resistance
on the domain walls in strained ferromagnetic manganite films
provide another interesting example of such properties that may become
important for various applications \cite{Golosov}. Last but not
least is a richness of magnetic and transport
properties of manganites. Concentration dependence and even overall topology pattern of
the phase diagram varies from material to material in these
compounds with general formula A$_{1-x}$B$_x$MnO$_3$ (A=La, Pr,
Nd; B=Sr, Ca) \cite{Tokura2}. In what follows we concentrate our
attention mainly on La$_{1-x}$Sr$_x$MnO$_3$ (LSMO) compounds
because there are no complications due to "charge ordering".
Instead their phase diagram is remarkable for the metallic
ferromagnetic and metallic anti-ferromagnetic phases inside the
concentration ranges $(0.3\le{x}\le{0.5})$ and
$(0.5\le{x}\le{0.7})$ correspondingly \cite{Tokura2}.
In addition, the LSMO compounds allow the use of band
description for broad concentration range.

Usually among the main difficulties in trying to describe the
properties of magnetic hetero-structures on the theory part lies
the fact that a contact area between the two materials with
different magnetic ordering corresponds to an abrupt change in the
underlying ground states: microscopic effects at the boundaries,
values of transmission coefficients of carriers, change in the
exchange mechanisms between the core spins {\it etc.}, all of them
taking place at the contact on the atomic scale. We are motivated by
the idea that the boundary between two manganite phases with rather
close doping levels may present a better ground for the understanding of interface
phenomena. It is worth mentioning, that attempts to fabricate and
investigate the characteristics of manganite hetero-structures
have already been made \cite{Izumi, Tanaka}.

In this paper we pursue a study of an electronic spectrum,
mechanisms of conductivity and magneto-conductivity of both the
ferromagnetic and anti-ferromagnetic (A-phase) phases. In
particular, we address the phenomena that may appear at the
interface of their contact. We use our results as an attempt to
rationalize some experimental observations \cite{Izumi} performed
on films and hetero-structures made of manganites having these two
phases as ground states.

Manganites are thought to belong to the group of the materials
with strongly correlated electrons. As it was mentioned above,
their phase diagram consists of numerous phases, with a crossover
from one phase to another upon change in temperature $T$, doping
concentration $x$ or composition, A$_{1-x}$B$_x$MnO$_3$, (see, for
example \cite{Tokura, Urushibara, Fujishiro} and references
therein). The parent LaMnO$_3$ compound has the
anti-ferromagnetic insulating state (A$^{\prime}$) at low temperatures.
Upon Sr substitution with La above $x>x_{cr}=0.16\div{0.17}$,
it undergoes a transition to a ferromagnetic (FM) metallic state.
With further increase in
doping ($x\geq{0.5}$), as it was mentioned above, there is a
transition from a ferromagnetic metallic to another
anti-ferromagnetic metallic state (A-phase). The latter is defined
as the phase, in which the core ($t_{2g}$) spins are aligned
ferromagnetically in the planes (for example, $ab$-plane) and
anti-ferromagnetically along the axis perpendicular to the planes
($c$-axis). In the ferromagnetic phase electrons are fully
polarized ("half-metallic" state). The "half-metallic" state is
also realized inside the $ab$-planes of the A-phase.
Below we restrict our consideration to low temperatures.

Despite of the mutual consent on the prevailing role of the double
exchange (DE) mechanism \cite{Zener} as the basic mechanism
responsible for ferromagnetism in manganites, there are several
theories describing their low-temperature properties from
different perspectives (for a recent review, see \cite{Dagotto}).
For our purposes it is worth mentioning right here that for the
concentration range ${0.3}\leq{x}\leq{0.75}$, the physical
properties of the LSMO compounds at least at low temperatures can
be accounted for within the double exchange (DE) model generalized
for the degenerate orbitals \cite{GK,Khomskii}. The scheme suggested
in \cite{GK,Khomskii,Solovyev} is basically the band model
utilizing however such important features of manganites as large
values of Hund's coupling constant, $J_H$, and double degeneracy
of the $e_{g}$ orbitals. It was shown to be capable of some
predictions regarding the evolution of the physical characteristics
with change in electron concentration upon doping. At larger
concentrations the super-exchange interaction between the $t_{2g}$
core spins also plays an important role \cite{Khomskii}.

The fact that the band picture captures the main physics of
manganites is not obvious. Often a generalized Hubbard model is
used to account for strong electron-electron interactions
\cite{Kugel}. However, instead of a direct Hubbard on-site $U>0$
which hinders the double occupation on each Mn site, one may
consider the local Jahn-Teller effect as 
another way to describe the same physics. Indeed,  a single
electron positioned on the degenerate $e_g$-orbital of the Mn site
will cause a local lattice distortion, reducing the energy
of a system. On the other hand two electrons on the same site do not lead to
Jahn-Teller instability and the Jahn-Teller energy gain does not
realize itself. Therefore in the local picture it is energetically favorable 
to have a single electron on a given site. We should also mention the large values of the
Hund's coupling ($J_H\simeq 1$eV), which is responsible for the
polarization of all electrons on the Mn site. A tendency
to the Jahn-Teller effect causes strong electron-phonon
interaction which is thought to be the reason for various
structural transitions in manganites, affecting their electronic
properties.

The Jahn-Teller instability is inherent to the \emph{local} high-spin
$(S=2)$ configuration of a manganese ion due to double degeneracy
of the $e_g$ orbitals. When such an instability realizes itself as
a \emph{cooperative} Jahn-Teller transition,  it also
accounts for the Coulomb interactions in a system, as it was
discussed above. In addition the coherent long-range order
correlations of the local distortions along the crystal lattice
favor the band description for electrons. Stability of the
anti-ferromagnetic (A'-phase) for the parent LaMnO$_3$ compound
was indeed explained in Ref. [\onlinecite{GK}]
in terms of the Jahn-Teller cooperative effect
incorporated into the two-band tight-binding approximation.

A primary manifestation of the Jahn-Teller effect comes about with
an appearance of a proper deformation of the MnO$_6$ octahedron.
However substitutional doping also leads to the lattice
deformations. Mismatch between the sizes of different ions is
expressed by the value of the tolerance factor, $t$:
\begin{equation}
t = \frac{1}{\sqrt{2}}\cdot\frac{R_B + R_O}{R_A + R_O},
\nonumber
\end{equation}
where $R_i (i=A, B, O)$ are the ionic radii of each element in
A$_{1-x}$B$_x$MnO$_3$. A process of doping results in an average
change of the lattice parameters, which may be described by the
average $\langle{t}\rangle_{av}$. When the value of
$\langle{t}\rangle_{av}$ is close to 1, the "cubic" perovskite
structure is realized as a whole. One must distinguish
$\langle{t}\rangle_{av}$ from its "local" value: two materials may
have close $\langle{t}\rangle_{av}$, i.e. close lattice
parameters, but the local distortions or disorder may differ
strongly from one site to another. Experiments show (see
\cite{Tokura} and references there in) that two compounds
with close overall $t\simeq{1}$ often have very different values
of  residual resistivity, $\rho_0$. Analysis in terms of the band
electrons is applicable if local disorder remains reasonably
small, i.e. the mean free path of electrons is large enough. From
this point of view, the considerable differences in the values of
conductivity of various ferromagnetic manganites at the same
doping level should be ascribed to strong variations in the values of the
local tolerance factor. With the increase in disorder the mean free path
may become so short that the Anderson mobility edge would hinder
metallic conductivity even in the ferromagnetic phase. Such
an unifying view turns out to be rather helpful for classification of
the physical properties of the whole series of doped manganites
\cite{GK}.

Before proceeding further another simple observation regarding the
role of the Jahn-Teller instability can be made. At $x=0$, according to
\cite{GK}, the A'-phase in LaMnO$_3$ is in the cooperative
state with all bands filled up by electrons. The Sr doping lifts the
local degeneracy of the $e_g$ orbitals and introduces random strains in the
lattice. De-localization in the electronic bands and hence
gain in the kinetic energy also competes with \emph{cooperative}
Jahn-Teller effects. Therefore, it is expected that at intermediate
concentrations $x$ the A'- phase is destroyed resulting in an appearance of
the ferromagnetic phase in the framework of the DE mechanism. 
There is no electron-hole symmetry in the two orbital model.
At $x=1$ there are no electrons in the system and,
therefore, departing from the ``other end'' of the phase diagram,
$x\to{1}$, small number of carriers, $1-x$, experience the Jahn-Teller 
instability and should be trapped locally at the Mn sites. 
With an increase in the number of
electrons, local Jahn-Teller traps will merge into a coherent state
which is responsible for another lattice structure with the directions of spins
remaining to be arranged. Although the
cooperative Jahn-Teller effect seems to be a major mechanism responsible
for the diversity of the phase diagram of manganites at
$x\simeq{0.5}$ and higher, spin interactions must be included to finally specify
the resulting ground states \cite{Dagotto,Khomskii}.

Due to the small differences in ionic radii between the La and Sr
atoms ($t\simeq{1}$), it seems, that the band description works
best for La$_{1-x}$Sr$_x$MnO$_3$ compounds \cite{GK}. For these
materials, being in the ferromagnetic metallic state, the residual
resistivity can be as small as $\rho_0\simeq{10^{-5}\div{10^{-4}}}
~(\Omega\cdot{cm})$~\cite{Quijada}, which is a good metallic
conductivity range. Unfortunately, so far there are not so many
experimental papers that deal with LSMO. Nevertheless, there are
few \cite{Izumi,Quijada,Urushibara, LSMO1, LSMO2} and in what follows we keep LSMO
materials in mind in our approach to different phenomena.

In this paper we first study the magneto-transport properties of
the manganites for a more general case of
a \emph{canted} A-phase \cite{deGennes}. This term means that the core $t_{2g}$
spins of the Mn ions maintain both the FM and AFM components
(e.g. spins in the A-phase can be canted by applying an external magnetic field).
We first derive an expression for the energy spectrum of electrons for
the canted A-phase manganites (Section II) and then proceed
with the derivation of the formulas for the magneto-conductivity by
generalizing the diagrammatic ``cross'' technique \cite{AGD} for
Green functions in the two-band model (Section III).
Specific case of manganites makes a modification in the form of the ``impurity potentials'' necessary,
to include a dependence on the orbital indexes. More specifically, the distortions
of the Mn-O-Mn conduction network are primarily caused by random octahedral distortions.
With Sr doping, the number of distorted octahedron increases so that one may expect a correlation
between the number of carriers and the value of conductivity even for the high-quality samples.

Throughout this paper we use the tight binding approximation in the frame of the two-orbital
DE model \cite{GK}. This model, as we already mentioned, has proved to capture the main
physical properties of manganites at least in the metallic concentration range. Finally in Section
IV we briefly discuss phenomena in the vicinity of the interface between the A-phase and FM-phase manganites.
The details of our numerical calculations with some exact results are provided in Appendix.
In the concluding Section together with a general discussion an attempt also
made to apply the results to recent experiments on
La$_{0.45}$Sr$_{0.55}$MnO$_3$/La$_{0.6}$Sr$_{0.4}$MnO$_3$
superlattices \cite{Izumi}.

\section{Energy spectrum of canted A-phase}
We start directly with a general case of the canted A-phase magnetic
structure. The band Hamiltonian has the form:
\begin{eqnarray}
\widehat{H}&=&\sum\limits_\mathbf{p}^{}
T^{\alpha\beta}({\mbox{\boldmath{$p$}}})~\widehat{a}^{\dagger}_{\alpha\sigma}
({\mbox{\boldmath $p$}})~\widehat{a}_{\beta\sigma}({\mbox{\boldmath $p$}}) +
J_H \sum\limits_\mathbf{p,Q}^{}
S({\mbox{\boldmath $Q$}})
~\widehat{a}^{\dagger}_{\alpha\sigma^{'}}({\mbox{\boldmath $p$}})
(\widehat{\sigma}_z)_{\sigma^{'}\sigma^{''}}
~\widehat{a}_{\alpha\sigma^{''}}({\mbox{\boldmath $p-Q$}}) + \nonumber\\
&&J_H \sum\limits_\mathbf{p,Q}^{}
S({\mbox{\boldmath $-Q$}})
~\widehat{a}^{\dagger}_{\alpha\sigma^{'}}({\mbox{\boldmath $p$}})
(\widehat{\sigma}_z)_{\sigma^{'}\sigma^{''}}
~\widehat{a}_{\alpha\sigma^{''}}({\mbox{\boldmath $p+Q$}}) +
J_H M\sum\limits_\mathbf{p}^{}
~\widehat{a}^{\dagger}_{\alpha\sigma^{'}}({\mbox{\boldmath $p$}})
(\widehat{\sigma}_x)_{\sigma^{'}\sigma^{''}}
~\widehat{a}_{\alpha\sigma^{''}}({\mbox{\boldmath $p$}}).
\label{eq1}
\end{eqnarray}
Here $T^{\alpha\beta}({\mbox{\boldmath $p$}})$ is an electron
hopping matrix for the two-band model, $J_H$ is the Hund's
coupling constant on the Mn sites, and $S({\mbox{\boldmath $Q$}})$
is the Fourier component of the AF ordering along the {\it c}
direction, $S_z(i)=\langle{S_z}\rangle{(-1)^i}$. The magnetic structural
vector ${\mbox{\boldmath $Q$}} = (0,0,\pi/a)$ reduces Brillouin
zone ($a$ is the cubic lattice constant); 
$M$ is a canted magnetic moment, so that at each site $i$:
\begin{equation}
{\mbox{\boldmath $S$}}(i)=(M_x, \pm\langle{S_z}\rangle), ~S_z^2 + M_x^2 \simeq{S^2}
\nonumber
\end{equation}
(when $S_z({\mbox{\boldmath $Q$}})=0$ we obtain the ferromagnetic phase).
The orientations of $M$ and $S_z$ are fixed by
magnetic anisotropy (easy plane) and/or by an external field. The
matrix elements $T^{\alpha\beta}({\mbox{\boldmath $p$}})$ in the
Eq.(\ref{eq1}) are calculated with the basis functions of the
following form \cite{GK}:
\begin{equation}
\psi_1\propto{z^2} + e^{i2\pi/3} x^2 + e^{-i2\pi/3} y^2,
\psi_2 = \psi_1^{*},
\label{eq2}
\end{equation}
As the result the hopping matrix elements are equal to:
\begin{equation}
T^{\alpha\beta}({\mbox{\boldmath $p$}}) = |A|
\left(\begin{array}{cc}
{T^{11}({\mbox{\boldmath $p$}})} &
{T^{12}({\mbox{\boldmath $p$}})}\\
{T^{21}({\mbox{\boldmath $p$}})} &
{T^{22}({\mbox{\boldmath $p$}})}
\end{array}\right),
\label{eq3}
\end{equation}
where
\begin{equation}
\begin{split}
T^{11}({\mbox{\boldmath $p$}}) = {T^{22}}({\mbox{\boldmath $p$}})
= {\cos(p_xa) + \cos(p_ya) + \cos(p_za)}, \\
T^{12}({\mbox{\boldmath $p$}})  = \left(T^{21}\right)^{*}({\mbox{\boldmath $p$}}) =
{\cos(p_za) + e^{i2\pi/3}\cos(p_xa) + e^{-i2\pi/3}\cos(p_ya)}
\end{split}
\nonumber
\end{equation}
and $|A|\simeq{0.16}$eV being a hopping amplitude \cite{GK}.

From (\ref{eq1}) we obtain the following
equation of motion:
\begin{equation}
(E\delta_{\alpha\beta} - T^{\alpha\beta}({\mbox{\boldmath
$p$}}))~\widehat{a}_{\beta\sigma} ({\mbox{\boldmath $p$}}) = J_H
S({\mbox{\boldmath $Q$}})
(\widehat{\sigma}_z)_{\sigma,\sigma^{'}}\widehat{a}_{\alpha\sigma^{'}}
({\mbox{\boldmath $p-Q$}}) + J_H
M(\widehat{\sigma}_x)_{\sigma,\sigma^{'}}
\widehat{a}_{\alpha\sigma^{'}}({\mbox{\boldmath $p$}})
\label{eq4}
\end{equation}
and similar equation for ${\mbox{\boldmath $p$}}\to
{\mbox{\boldmath $p+Q$}}$. Thus, the secular equation,
is now an $8\times{8}$ determinant from which one must calculate
the eigenvectors and eigenvalues. Let us remind that the
double exchange (DE) mechanism for the manganites \cite{Zener}
exploits the large value of the Hund interaction, $J_H\approx{1\div{2}}$
eV, so that $J_H/|A|\gg{1}$. Using this approximation, we solve Eq.
(\ref{eq4}) up to the terms of the order of $|A|^2/J_H$. The
electrons can occupy only four lowest bands:
\begin{equation}
\begin{split}
E_{1,2}({\mbox{\boldmath $p$}};M/S) = -J_HS - |A|{\cdot}[c_x + c_y +
(M/S)c_z \pm R_{12}({\mbox{\boldmath $p$}};M/S)], \\
E_{3,4}({\mbox{\boldmath $p$}};M/S) = -J_HS - |A|{\cdot}[c_x + c_y -
(M/S)c_z \pm R_{34}({\mbox{\boldmath $p$}};M/S)],
\end{split}
\label{eq5}
\end{equation}
where for brevity we introduced the notations
$c_i\equiv{\cos(p_ia)}, i=x,y,z$ and
\begin{equation}
R_{12}({\mbox{\boldmath $p$}}; M/S) =
\sqrt{c_x^2 + c_y^2 + (M/S)^2 c_z^2 - (M/S)c_z(c_x+c_y)
-c_x c_y}, ~R_{34}({\mbox{\boldmath $p$}}; M/S) =
R_{12}({\mbox{\boldmath $p$}}; -M/S).
\nonumber
\end{equation}
Performing the canonical transformation in accordance with the
eigenvalues of Eqs. (\ref{eq5}), it is straightforward to express
operators $\widehat{a}^{\dagger}_{\alpha\sigma}({\mbox{\boldmath
$p$}})$ and $\widehat{a}_{\alpha\sigma}({\mbox{\boldmath $p$}})$
in Eq. (\ref{eq1}) in terms of the new eigenfunctions. Quite
generally, the transformation has the form:
\begin{equation}
\widehat{a}_{\alpha\sigma}({\mbox{\boldmath $p$}}) =
\sum\limits_{l=1}^{8} K^{(l)}_{\alpha\sigma}
({\mbox{\boldmath $p$}})\cdot\widehat{\zeta}_l({\mbox{\boldmath $p$}}),
\label{eq6}
\end{equation}
where $\widehat{\zeta}^{\dagger}_l({\mbox{\boldmath $p$}}),
\widehat{\zeta}_l({\mbox{\boldmath $p$}})$ are creation and annihilation
operators for the true energy branches (\ref{eq5}),
$\sigma\equiv(\uparrow\downarrow)$. Below we write down explicitly
the expressions for $K^{(l)}_{\alpha\sigma}$ for the four lowest
bands:
\begin{equation}
\begin{split}
K^{(1)}_{1\sigma}({\mbox{\boldmath $p$}}) =
K^{(2)}_{1\sigma}({\mbox{\boldmath $p$}}) =
\frac{1}{2}(1+M/S)^{1/2}\left(\frac{\Sigma_{12}({\mbox{\boldmath
$p$}})}{2R_{12}({\mbox{\boldmath $p$}})}\right)^{1/2},
~K^{(1)}_{2\sigma}({\mbox{\boldmath $p$}}) =
-K^{(2)}_{2\sigma}({\mbox{\boldmath $p$}}) =
\frac{1}{2}(1+M/S)^{1/2}\left(\frac{\Sigma^{*}_{12}({\mbox{\boldmath $p$}})}
{2R_{12}({\mbox{\boldmath $p$}})}\right)^{1/2},\\
K^{(3,4)}_{1\uparrow}({\mbox{\boldmath $p$}}) = -
K^{(3,4)}_{1\downarrow}({\mbox{\boldmath $p$}}) =
\frac{1}{2}(1-M/S)^{1/2}\left(\frac{\Sigma_{34}({\mbox{\boldmath
$p$}})}{2R_{34}({\mbox{\boldmath $p$}})}\right)^{1/2},
K^{(3,4)}_{2\uparrow}({\mbox{\boldmath $p$}}) = -
K^{(3,4)}_{2\downarrow}({\mbox{\boldmath $p$}}) =
\pm\frac{1}{2}(1-M/S)^{1/2}\left(\frac{\Sigma^{*}_{34}({\mbox{\boldmath
$p$}})}{2R_{34}({\mbox{\boldmath $p$}})}\right)^{1/2},
\end{split}
\label{eq7}
\end{equation}
where the following notation has been used:
\begin{equation}
\begin{split}
\Sigma_{12}({\mbox{\boldmath $p$}}) = (M/S)c_z - \frac{1}{2}(c_x +
c_y) + i\frac{\sqrt{3}}{2}(c_y - c_x), \\
~\Sigma_{34}({\mbox{\boldmath $p$}}) = -(M/S)c_z - \frac{1}{2}(c_x
+ c_y) + i\frac{\sqrt{3}}{2}(c_y - c_x).
\end{split}
\nonumber
\end{equation}
\section{Conductivity and magneto-conductivity of canted A-phase}
It is straightforward to extend the standard ``cross'' technique
for static defects \cite{AGD} for the two-band model. As for the nature of the defects,
when an ion B is substituted for an ion A in the unit formula
A$_{1-x}$B$_x$MnO$_3$ it immediately lifts the cubic symmetry at
the Mn-site. The $e_g$ doublet gets split and the oxygen octahedron
becomes distorted. Since, in accordance with our introductory
remarks, this effect is of prime importance for the Mn-O-Mn
conduction network, disorder in manganites to a large extent
comes about through a change in doping. In application of diagrammatic
cross-technique, however, we assume that positions of
``impurities'' (i.e. of B ions) remain random and there is no
correlations between the scattering processes from any two of
them. Secondly, for the ``impurity'' potential
$U_{\alpha\beta}({\mbox{\boldmath $r-R_i$}})$ in
\begin{equation}
\widehat{H}_{imp} = \sum\limits_{i} \int{d^3\mbox{\boldmath $r$}}
{\psi}_{\alpha\sigma}^{\dagger}({\mbox{\boldmath $r$}})
U_{\alpha\beta}({\mbox{\boldmath $r-R_i$}})
{\psi}_{\beta\sigma}({\mbox{\boldmath $r$}}) \label{eq8}
\end{equation}
(where summation is held over the random realizations of the
``impurities'') one can assume the Jahn-Teller form of the defect potential.
Using the basis given by Eq. (\ref{eq2}), the
expression for $U_{\alpha\beta}$ is:
\begin{equation}
U_{\alpha\beta}({\mbox{\boldmath $r-R_i$}}) =
gQ({\mbox{\boldmath $R_i$}})\cdot
\left(\begin{array}{cc} 0 & e^{i\theta_i} \\
e^{-i\theta_i} & 0 \end{array}
\right)_{\alpha\beta}\cdot{\delta}({\mbox{\boldmath $r-R_i$}}),
\label{eq9}
\end{equation}
where $Q({\mbox{\boldmath $R_i$}})$ is an amplitude of the
Jahn-Teller distortion at site $i$, $g$ is an electron-lattice coupling constant and
the angle $\theta_i$ specifies the shape of the distorted
octahedron at a given Mn site. Going to momentum representation,
the expression for $\widehat{H}_{imp}$ is:
\begin{equation}
\widehat{H}_{imp} = \sum\limits_{i}
\int\int\frac{d^3{\mbox{\boldmath $p$}}}{(2\pi)^3}
\frac{d^3{\mbox{\boldmath $p'$}}}{(2\pi)^3}
\widehat{a}_{\alpha\sigma}^{\dagger}({\mbox{\boldmath $p$}})
U_{\alpha\beta}({\mbox{\boldmath $p-p'$}})
\widehat{a}_{\beta\sigma}({\mbox{\boldmath $p'$}})
{\cdot}e^{i({\mbox{\boldmath $p-p'$}})\cdot{\mbox{\boldmath
$R_i$}}}, \tag{9a} \label{eq9a}
\end{equation}
with
\begin{equation}
U_{\alpha\beta}({\mbox{\boldmath $p$}}) = \int{d^3\mbox{\boldmath
$r$}} U_{\alpha\beta}({\mbox{\boldmath $r$}})
\cdot{e}^{-i{\mbox{\boldmath $p\cdot r$}}}. \nonumber
\end{equation}
Keeping in mind the energy spectrum obtained in the preceding
Section it is helpful to re-write expression (\ref{eq9a}) in terms
of new variables defined by (\ref{eq6}):
\begin{equation}
\begin{split}
\widehat{H}_{imp} = \sum\limits_{i}gQ({\mbox{\boldmath $R_i$}})
\int\int\frac{d^3{\mbox{\boldmath $p$}}}{(2\pi)^3}
\frac{d^3{\mbox{\boldmath $p'$}}}{(2\pi)^3}
\sum\limits_{l_1,l_2=1}^{4}{~\mathcal M}^{(l_1l_2)}
({\mbox{\boldmath $p,p'$}};i)
\cdot\widehat{\zeta}_{l_1}^{\dagger}({\mbox{\boldmath
$p$}})\cdot~u({\mbox{\boldmath
$p-p'$}})\cdot~\widehat{\zeta}_{l_2}({\mbox{\boldmath $p'$}})
e^{i({\mbox{\boldmath $p-p'$}})\cdot{\mbox{\boldmath $R_i$}}},\\
{\mathcal M}^{(l_1l_2)}({\mbox{\boldmath $p,p'$}};i)\equiv
\sum\limits_{\sigma={(\uparrow\downarrow)}}
\left\{K_{1\sigma}^{(l_1)*}({\mbox{\boldmath $p$}}) \cdot
K_{2\sigma}^{(l_2)}({\mbox{\boldmath $p'$}}) ~e^{i\theta_i} +
K_{2\sigma}^{(l_1)*}({\mbox{\boldmath $p$}})\cdot
K_{1\sigma}^{(l_2)}({\mbox{\boldmath
$p'$}})~e^{-i\theta_i}\right\}.
\end{split}
\tag{9b} \label{eq9b}
\end{equation}
The so-called ``cross-technique'' \cite{AGD} can now be
straightforwardly applied to the calculation of the average of new
band Green function given by:
\begin{equation}
G_{l}({\mbox{\boldmath $p,p'$}};t) = -i{\langle}\widehat{T}
\left\{\widehat{\zeta}_{l}({\mbox{\boldmath $p$}};0)
\widehat{\zeta}^{\dagger}_{l}({\mbox{\boldmath $p'$}};t)
\right\}\rangle \label{eq10}
\end{equation}
In the absence of the defects, the Green function (\ref{eq10}) is:
\begin{equation}
G^{(0)}_{l}({\mbox{\boldmath $p$}};\varepsilon) =
\frac{1}{\varepsilon - \xi_l({\mbox{\boldmath $p$}}) +
i0~\mathrm{sign}\varepsilon}, \label{eq12}
\end{equation}
where $\xi_l({\mbox{\boldmath $p$}}) = E_l({\mbox{\boldmath $p$}})
- E_F$ (again we can leave only four essential bands, since
$J_H\gg{|A|}$).
\begin{figure}[h]
\centerline{\psfig{figure=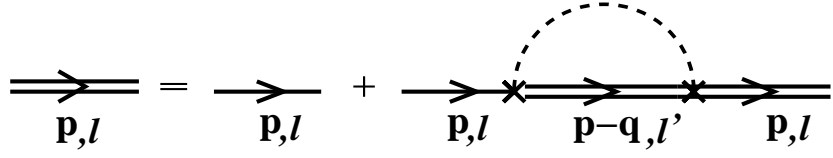,height=2cm,width=11cm,angle=0}}
\caption{Equation for an averaged over defects Green function
(\ref{eq10}). Summation over $l'$ and integration over ${\mbox{\boldmath $q$}}$ are assumed.}
\label{fig74}
\end{figure}
For the Green function averaged over defect's positions
\begin{equation}
\langle{G({\mbox{\boldmath $p,p'$}};\varepsilon)}\rangle_{dis} =
{G({\mbox{\boldmath
$p$}};\varepsilon)}\cdot\delta({\mbox{\boldmath $p-p'$}})
\label{eq13}
\end{equation}
we obtain the well known form of equation schematically shown in
Fig.\ref{fig74}. The self-energy part
$\widehat{\Sigma}_l({\mbox{\boldmath $p-q$}};\varepsilon)$ on
Fig.\ref{fig74} is again expressed in terms of the corresponding
relaxation times as:
\begin{equation}
\widehat{\Sigma}_l(\varepsilon) = -i~\mathrm{sign}\varepsilon
\left\langle\frac{\hbar}
{2\tau_l({\mbox{\boldmath $p$}})}\right\rangle_{F.S.},
\label{eq14}
\end{equation}
where $\langle{...}\rangle_{F.S.}$ denotes an average over the
Fermi surface. Let us emphasize that attenuation $\tau_l$ in (\ref{eq14})
contains contributions from scattering between different bands (\ref{eq5}).
In the representation (\ref{eq9b}), the
expressions for relaxation times ${\hbar}/{2\tau_l}$ are:
\begin{equation}
\begin{split}
\frac{\hbar}{2\tau_l} = \pi\frac{|gQ|^2}{2\nu_l(E_F)}n_{imp}
\int\frac{d^3{\mbox{\boldmath $p$}}}{(2\pi)^3}
\int\frac{d^3{\mbox{\boldmath
$p'$}}}{(2\pi)^3}\sum\limits_{l_1=1}^{4}~M^{(ll_1)}({\mbox{\boldmath
$p,p'$}}){\cdot}\delta(E_F - E_{l_1}({\mbox{\boldmath
$p'$}}))\cdot M^{(l_1l)}({\mbox{\boldmath $p',p$}}) \delta(E_F -
E_l({\mbox{\boldmath $p$}})),\\
M^{(l_1l_2)}({\mbox{\boldmath
$p,p'$}})\equiv\sum\limits_{\sigma=(\uparrow,\downarrow)}^{}
\left\{K_{1\sigma}^{(l_1)*}({\mbox{\boldmath $p$}}) \cdot
K_{2\sigma}^{(l_2)}({\mbox{\boldmath $p'$}}) +
K_{2\sigma}^{(l_1)*}({\mbox{\boldmath $p$}})\cdot
K_{1\sigma}^{(l_2)}({\mbox{\boldmath $p'$}})\right\}
\end{split}
\label{eq15}
\end{equation}
where $n_{imp}$ is concentration of "impurities" (if our model with the potential (\ref{eq8}-\ref{eq9})
for Sr atom, one would have for $n_{imp} = x$, where $x$ is a dopant concentration; we will come back 
to this in Section IV);
$\nu_l(E_F)$ is a density of
states at the Fermi level of the $l$th band. In the process of derivation of Eq.
(\ref{eq15}) we took into account that the main contribution to
the integrals in Eq. (\ref{eq15}) comes from the region close to
the Fermi surface. We also used the following averages over
disorder: ${\langle\left({gQ}({\mbox{\boldmath
$R$}}_i)\right)^2\rangle_{dis}} = |gQ|^2$,
$\langle{e}^{2i\theta_i}\rangle_{dis} = 1$ (the latter one means averaging over the local
shapes of distorted octahedra).

With the use of Eqs. (\ref{eq14},\ref{eq15}), the expression
for the averaged Green function (\ref{eq13}) can be
written in the form:
\begin{equation}
\langle{G_{l} ({\mbox{\boldmath
$p,p'$}};\varepsilon)}\rangle_{dis} = \delta({\mbox{\boldmath
$p-p'$}})\cdot\frac{1}{\varepsilon - \xi_l({\mbox{\boldmath $p$}})
+ i\left({\hbar}/{2\tau_l}\right)\mathrm{sign}\varepsilon}.
\label{eq16}
\end{equation}

The dc-conductivity can be calculated from the Kubo formula (see, for example, in [\onlinecite{Doniach}]):
\begin{equation}
\sigma_{\alpha\alpha}(0) = \lim\limits_{\omega\to{0}}^{}
\frac{R_{\alpha\alpha}(\omega)}{i\omega}. \label{eq17}
\end{equation}
where $R_{\alpha\alpha}(\omega) (\alpha{=}x,y,z)$ can be obtained with the help
of the corresponding product of retarded Green functions, averaged over impurities:
\begin{equation}
\langle{R_{\alpha\alpha}(\omega)}\rangle_{dis}=-\frac{ie^2\hbar}{(2\pi)}
\int\limits_{-\infty}^{\infty}d\varepsilon\sum\limits_{l}^{}
\int\int\frac{d^3{\mbox{\boldmath $p$}}}{(2\pi)^3}
\frac{d^3{\mbox{\boldmath $p'$}}}{(2\pi)^3}
\left\langle\left({\mbox{\boldmath $\widehat{v}$}}^{(l)}_\alpha
{\widehat{G}_{l}({\mbox{\boldmath
$p,p'$}};\varepsilon+\omega)}\right) \cdot\left({\widehat{G}_{l}
({\mbox{\boldmath $p',p$}};\varepsilon)} {\mbox{\boldmath
$\widehat{v'}$}}^{(l)}_\alpha\right)\right\rangle_{dis} \label{eq18}
\end{equation}
via analytic continuation $R_{\alpha\alpha}(i\omega_n){\to} R_{\alpha\alpha}(\omega +
i\delta)$ (in Eq. (\ref{eq18}) ${\mbox{\boldmath$\widehat{v}$}}^{(l)}_\alpha$
is the velocity operator defined as a derivative of energy with
respect to the momentum for each band given by Eqs. (\ref{eq5})). With
the impurity potential given by Eqs. (\ref{eq9}, \ref{eq9a}),
the average of product in expression (\ref{eq18}) can be
re-written as
\begin{equation}
\left\langle{\mbox{\boldmath
$\widehat{v}$}}^{(l)}_\alpha{\widehat{G}_{l} ({\mbox{\boldmath
$p,p'$}};\varepsilon+\omega)} {\cdot}
\widehat{G}_{l}({\mbox{\boldmath $p',p$}};\varepsilon)
{\mbox{\boldmath $\widehat{v'}$}}^{(l)}_\alpha\right\rangle_{dis}
= {\mbox{\boldmath $\widehat{v}$}}^{(l)}_\alpha
\left\langle{\widehat{G}_{l} ({\mbox{\boldmath
$p,p'$}};\varepsilon+\omega)} \right\rangle_{dis}\cdot
\left\langle{\widehat{G}_{l} ({\mbox{\boldmath
$p',p$}};\varepsilon)}\right\rangle_{dis} {\mbox{\boldmath
$\widehat{v'}$}}^{(l)}_\alpha. \label{eq19}
\end{equation}
Now, taking into account equations (\ref{eq13}-\ref{eq19}) and
performing the integration in (\ref{eq18}) with respect to
$\varepsilon$, we finally obtain the following expression for the
in-plane and out-of-plane dc-conductivities:
\begin{eqnarray}
\sigma_{xx} &=&\sigma^{(+)}_{xx} + \sigma^{(-)}_{xx},\nonumber\\
\sigma_{zz} &=&\sigma^{(+)}_{zz} + \sigma^{(-)}_{zz},
\nonumber
\end{eqnarray}
where
\begin{equation}
\begin{split}
\sigma^{(+)}_{\alpha\alpha} = (1+M/S)^2\cdot\frac{e^2}{a\hbar}
\int\limits_{F.S.}^{}\sum\limits_{l=1}^{2}
\frac{\tau_l}{\hbar}\cdot
\frac{dS^l_{\mbox{\boldmath $p$}}}{|\nabla_{\mbox{\boldmath $p$}}E_l|}
\left(\frac{\partial{E_l({\mbox{\boldmath $p$}})}}
{\partial{{\mbox{\boldmath $p$}}_\alpha}}\right)^2, \\
\sigma^{(-)}_{\alpha\alpha} = (1-M/S)^2\cdot\frac{e^2}{a\hbar}
\int\limits_{F.S.}^{}\sum\limits_{l=3}^{4}\frac{\tau_l}{\hbar}\cdot
\frac{dS^l_{\mbox{\boldmath $p$}}}{|\nabla_{\mbox{\boldmath
$p$}}E_l|} \left(\frac{\partial{E_l({\mbox{\boldmath $p$}})}}
{\partial{{\mbox{\boldmath $p$}}_\alpha}}\right)^2,
\end{split}
\label{eq20}
\end{equation}
where $\hbar/\tau_l (l=1,2,3,4)$ are defined by Eq. (\ref{eq15}), $\alpha=(x,z)$ and
integration runs over each Fermi surface.
\begin{figure}[h] \hspace{-1cm}
\centerline{\psfig{file=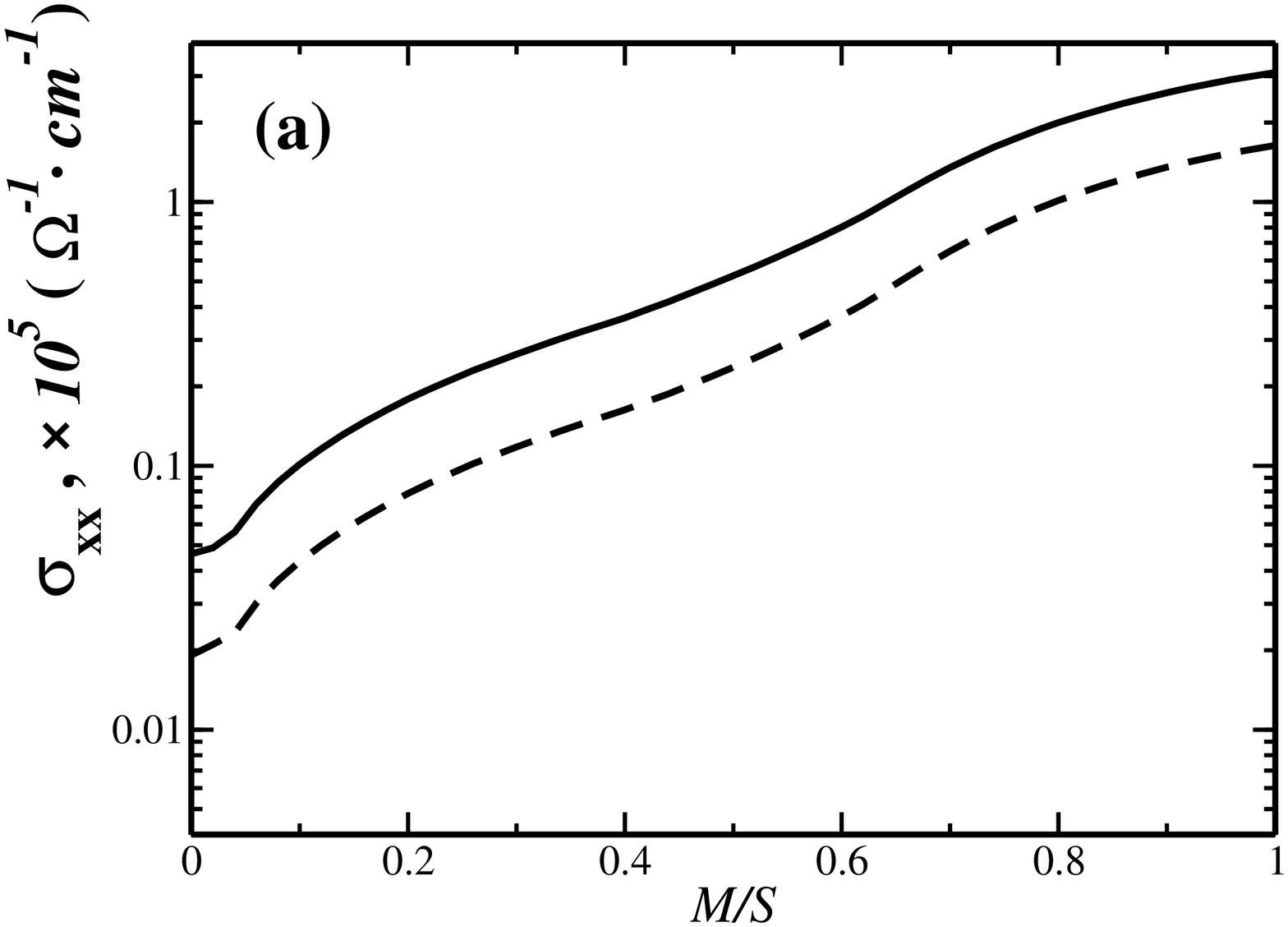,height=6cm,width=8cm,angle=-90}}
\centerline{\psfig{file=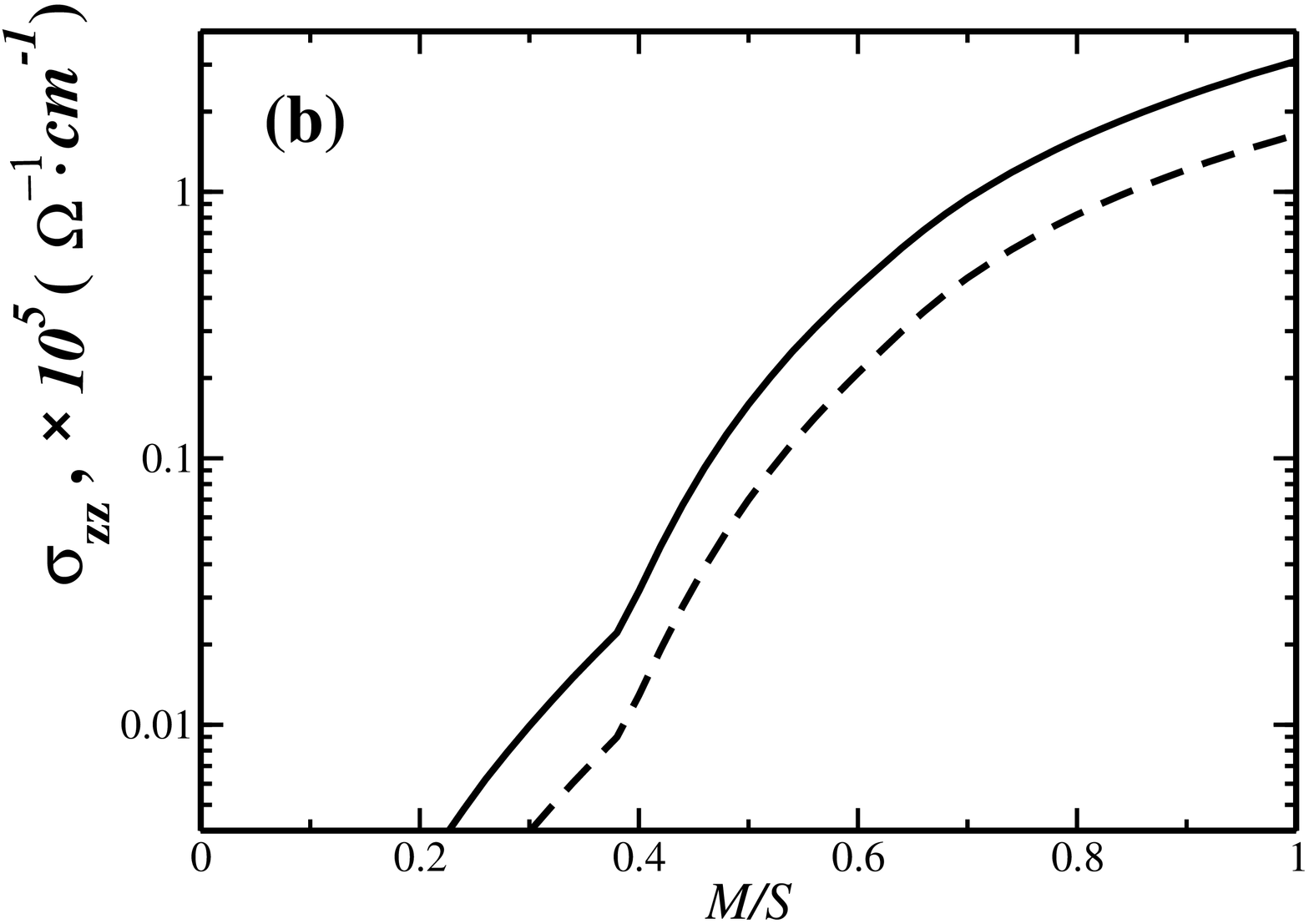,height=6cm,width=8cm,angle=-90}}
\caption{In-plane (a) and out-of-plane (b) conductivities in the canted A-phase 
are shown on $\log$ scale as a function of $M/S$ for different values 
of $\Delta=|gQ|/|A|$. Solid (dashed) lines correspond to the $\Delta = 0.45~(0.75)$. Both
curves are obtained for dopant concentration $x=0.55$ and $a=3.9\AA$. The
dependence $E_F(M/S)$ is presented below on Fig.\ref{MSvsEF}.} 
\label{cond}
\end{figure}

\begin{figure}[h]
\hspace{-1cm}
\centerline{\psfig{file=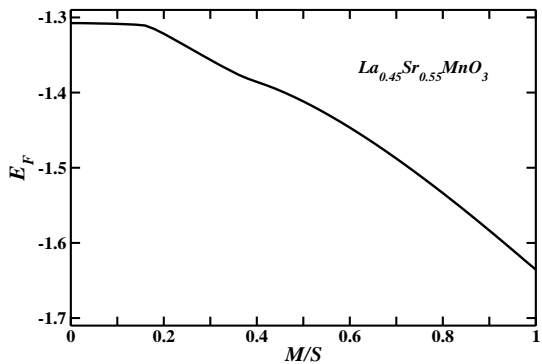,height=6cm,width=8cm,angle=-90}}
\caption{The dependence of $E_F$ on $M/S$ for a given
concentration $x=0.55$ based on band structure described by
(\ref{eq5}). $E_F$ is given in the units of the hopping amplitude
$|A|$ (see Eq. (\ref{eq5})). The reference point for the Fermi
level is taken at the bottom of the lowest band.} \label{MSvsEF}
\end{figure}

Expressions (\ref{eq20}) for both in-plane and out-of-plane
conductivities are rather complicated and the procedure of evaluating
the conductivity using them deserves further explanation.
Parameters of the electronic spectrum are chosen beginning with the initial 
two-band ``cubic'' phase which, as we believe, realizes itself in the
ferromagnetic state. Let us also recall that the disorder in the
calculations presented above expresses itself through the local
octahedra distortion which, in turn, is produced by the
substitutional disorder, i.e. by non-stoichiometric
A$_{1-x}$B$_x$MnO$_3$. Therefore, in (\ref{eq20}) the
"concentration" of "impurities" is the concentration of B atoms,
while the amplitude of the disorder potential is given by the
value of $|gQ|$ in Eq. (\ref{eq15}). Change in composition by
increasing $x$ decreases the number of carriers, $1-x$, while
increasing the number of defects \emph{simultaneously}. Another
significant simplification above was that the distortions of the oxygen
octahedra were treated independently. Indeed this is a good approximation, 
because two octahedra surrounding two neighboring
Mn atoms share one oxygen atom only. If it were not so, the values of
resistivity would depend on the B atom concentration only. In
reality a sample's quality also depends on how these octahedra
adjust themselves. Factor $n_{imp}$ in (\ref{eq15}) is not the
only reason for dependence of conductivity on $x$ in (\ref{eq20}).
Change in carrier concentration results in a shift of the
chemical potential relative to the bottom of the bands which
reflects itself in an immediate change in the occupation number
in each of the four active energy bands (\ref{eq5}). Such a
non-trivial intimate dependence between the number of carriers and
the number of defects presents itself as a new feature for
conductivity behavior in manganites. It would be of great interest to investigate
such a trend experimentally. Currently a shortage of
experimental data for LSMO compounds for large enough Sr
concentrations deprives us the possibility to trace that
dependence in some more details. Some estimates have been done for
the FM-phase in our previous paper \cite{GK} for $x=0.3\div{0.4}$.
In this presentation we perform the calculations for canted A-phase for
$x=0.55$. In Fig. \ref{cond}(a,b) we plotted our results. Energy
spectrum of the A-phase itself in the DE approximation
($J_H\gg{|A|}$) would not allow current to flow in the
out-of-plane direction: the dispersion $t(p_z)\propto\cos{(p_z)}$
drops out from (\ref{eq5}) for $M=0$. Therefore Fig. \ref{cond}
describes, as expected, a dramatic magneto-resistance effect
inherent to the canted A-phase for the perpendicular-to-the-plane
current ($\sigma_{zz}$). Surprisingly,  it turned out
that even the in-plane ($\sigma_{xx}$) components of conductivity
display considerable change in its value at the transition from
the 3D conductivity regime in the ferromagnetic state (i.e. at
$M/S=1$) to the 2D one for the A-phase ($M=0$). The origin of such
a rapid change comes about from the
re-distribution of carriers between the energy bands with change
in the value of $M/S$. The effect of carrier re-distribution in the bands is seen in Fig.
\ref{MSvsEF} which shows the calculated position of the Fermi
level for doping concentration $x=0.55$. With variation of $M/S$ the
system undergoes dimensional transition between the 2D and 3D
conductivity regimes.

In Fig. \ref{cond} (a,b) it seemed more convenient to present our
results for the chosen concentration as a function of $M/S$ in
accordance with Eqs. (\ref{eq19}). We remind that $M$ is a
ferromagnetic component of the \emph{core spins} only. In order to
find the values of conductivity as a function of the total
magnetization, $\widetilde{M}$, which also includes the electronic
component and is induced by an external magnetic field, one may use
the following simple relation:
\begin{equation}
\widetilde{M} = \mu_B{(4-x)}\cdot(M/S).
\label{eq21}
\end{equation}
Equation (\ref{eq21}) expresses the value of the full magnetic
moment $\widetilde{M}=\chi{B}$, in units of Bohr magneton per
Mn ion, and $\chi$ is a magnetic susceptibility.
\section{Charges and spins near interface}
We now turn to some problems which involve artificial contacts
of manganites with different magnetic ground states or just two
manganites having the same ground states but different doping
concentrations. We will discuss both charge and magnetic structure
in the vicinity of the boundary. Below we consider the plane
geometry, so that all quantities depend on one coordinate only. In
addition we simplify our discussion by choosing the single band DE
model $T^{\alpha\beta}\to{t}$, where $t$ is a hopping amplitude.

We first consider two ferromagnetic manganites with different
doping concentrations brought into contact with each other with
the parallel orientations of the local moments. Some pronounced
effects come about due to the Schottky layers formed at the
contact. Difference in doping concentrations produces a difference
in the chemical potentials on both sides away from the contact,
${E_F}^{left}$ and ${E_F}^{right}$, i.e. difference in the "work
functions" of two components. That leads to a redistribution of
carriers near the contact plane. This effect is general and well
known for the contacts between metals or semiconductors. A
simplifying feature for the contacts of two manganites is in
similarity of the underlying band structures on the both sides of
a contact. At the same time all major changes still take place at
the atomic scale so that one needs to apply the Kohn-Sham scheme
to solve for potential and charge distributions self-consistently.
To elucidate some qualitative features we provide in Appendix the
solution of this problem in its continuous formulation.

\begin{figure}[h]
\hspace{-1cm}
\centerline{\psfig{file=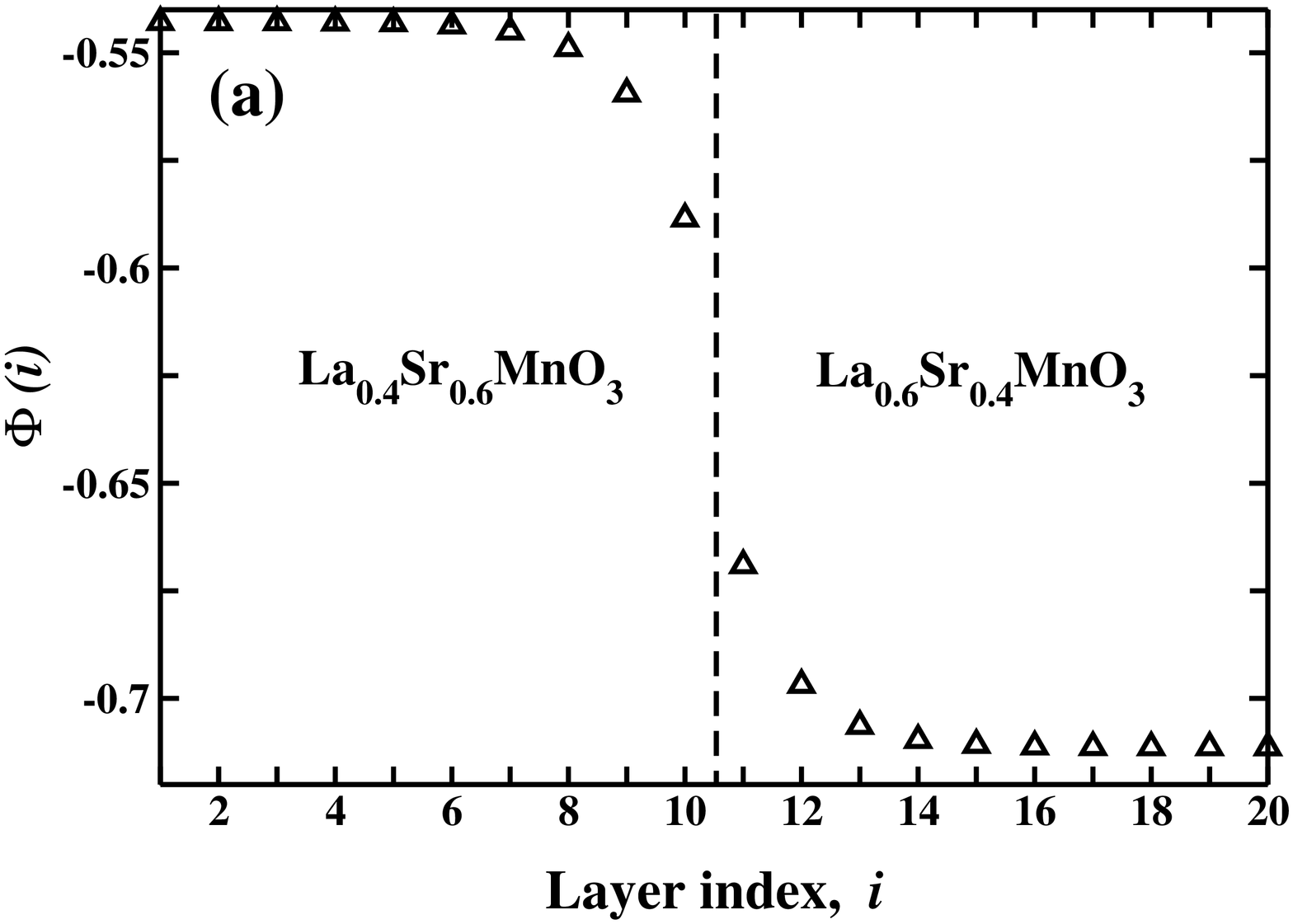,height=6cm,width=8cm,angle=-90}}
\centerline{\psfig{file=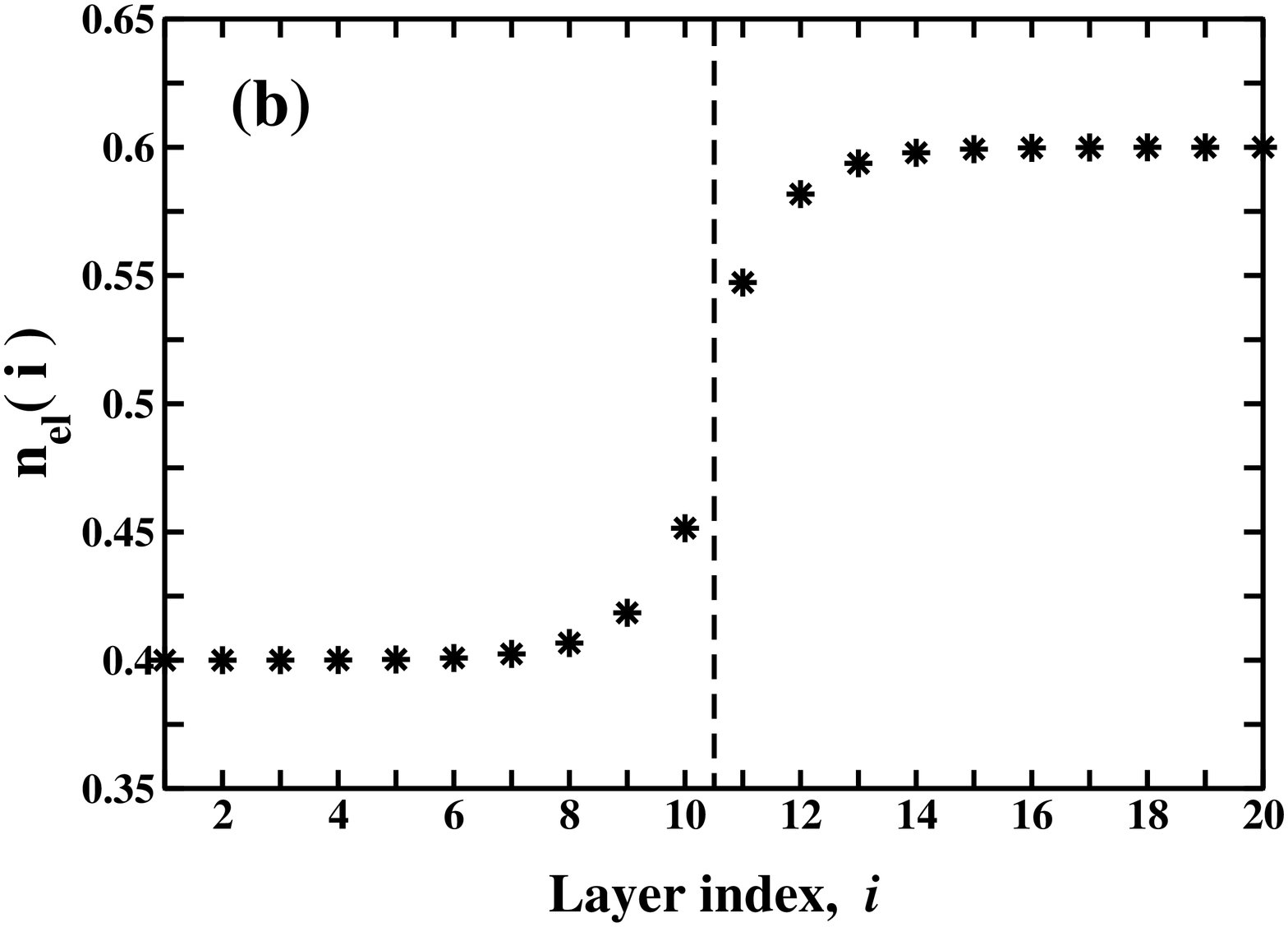,height=6cm,width=8cm,angle=-90}}
\caption{Contact of two metallic ferromagnetic phases with
different Sr concentrations in the one band DE model. The
numerical solution of Kohn-Sham equations is shown for the
structure with number of layers being equal to twenty. The dashed line
shows the position of the (La,Sr) plane, where the abrupt change in Sr
concentrations takes place; a) Potential $\Phi(i)$ is given in dimensionless
units $4\pi{e^2}/(at)$; b) $n_{el}(i)$ is concentration of electrons in the layer 
$i$ in units of $a^{-2}$, where $a$ is a lattice constant.} \label{PotEl}
\end{figure}

We proceed as follows. Let $N_{L(R)}$ be the concentration (i.e.
the number per cm$^2$) of positive charges in (LaSr) plane on the
left(right) side far away from the contact. In the process of
preparation of the hetero-structure (film deposition)\cite{Izumi},
the Sr concentration $N_{Sr}$ changes sharply from $N_L$ to $N_R$
at the contact. The system of Kohn-Sham equations
(the Poisson equation for the potential distribution and the
Schr\"{o}dinger equation) is:
\begin{equation}
\begin{split}
\Phi{(i+1)} - 2\Phi{(i)} + \Phi{(i-1)} = \frac{4\pi{e^2}}{t}\left[N_{Sr}(i) - n_{el}(i)\right],\\
-\Psi_{\lambda}{(i+1)} + 2\Psi_{\lambda}{(i)} -
\Psi_{\lambda}{(i-1)} + \Phi(i)\Psi_{\lambda}(i) =
E_{\lambda}\Psi_{\lambda}(i),
\end{split}
\label{h-1}
\end{equation}
where $i$ is an index, which runs through the Mn planes, $\lambda$
is an eigenvalue index, $E_\lambda$ is an energy in units of $t$,
$\Phi$ is a dimensionless potential defined by electrostatic
potential $\varphi(i)$ as
\begin{equation}
\Phi(i) = -\frac{|e|\varphi(i)}{t},
\label{h-1p}
\end{equation}
$n_{el}(i)=\sum\limits_{\lambda{<}\lambda_F}|\Psi_{\lambda}(i)|^2$
is a concentration of electrons on a plane $i$ and $N_{Sr}(i)$ is a Sr
concentration, which depends on which side of the contact an electron
is located and $E_{\lambda_F} = E_F(i)$ with $E_F(i)$ being equal to
the local Fermi level in the units of $t$. We have obtained a numerical solution of
(\ref{h-1}) with the boundary conditions providing the equality of the
electrochemical potentials across the contact. In the calculations
below the total number of layers was equal to twenty (ten on each
side of the contact) with:
\begin{equation}
N_{Sr}(i) = \left\{
\begin{array} {r@{\quad}}
0.6, ~i\le{10}, \\
0.4, ~i>{11}.
\end{array}
\right.
\label{Nsr}
\end{equation}
The solution for the potential and electron distribution is shown on
Fig. \ref{PotEl}(a,b).
The consistency of our numerical results have been verified by
comparing them with the analytical ones obtained from the solution
of the same problem in its continuous form. The latter are
described in Appendix in more details.

The significance of these calculations for the further discussion
is as follows. First of all, one sees that the electron screening
(the ``Thomas-Fermi'' length) extends over four-five atomic
distances implying that the sharpness of the contact is
smoothed out considerably. Secondly, there is a re-distribution of
charge between phases: repletion and depletion regions form close
to the interface.

We now turn to the discussion of a contact between the
manganites in the ferromagnetic and anti-ferromagnetic A-phase.
The results of the previous Section have demonstrated the
pronounced magneto-resistance effects in the canted A-phase for
both in-plane and out-of-plane directions of a current. This
justifies an interest and need for better  understanding of the
F/A contact properties.

Stacks of manganite films of different thickness with abrupt 
change in Sr concentration in the (LaSr) planes in order to stabilize FM- or A-phase, 
can be made using the state of the art deposition techniques \cite{Izumi}. 
However contacts between the FM state and the A-phase studied in \cite{Izumi} pose more
questions than the preceding example of the contact between the two
unequally doped ferromagnetic metallic manganites.
\begin{figure}
\hspace{0cm}
\centerline{\psfig{file=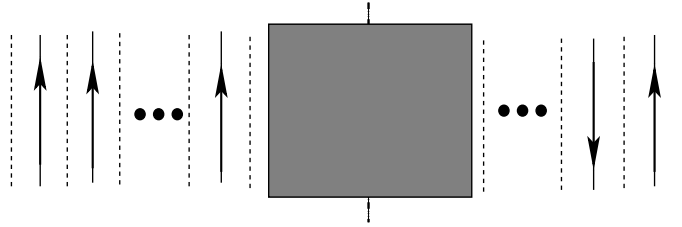,height=3cm,width=9cm,angle=0}}
\caption{Schematic presentation of a contact between ferromagnetic
(left) and A-phase (right) manganites. Shaded area designates the
interface region. 
The dashed lines designate the (LaSr) layers (the
long dashed line in the middle singles out the plane (interface layer) in which the
Sr concentration is sharply changed). Full lines designate the MnO
planes in which carriers are positioned. The directions of the
arrows in Fig. \ref{spins} to the left and to the right from the
interface show the magnetic order in the FM- and A-phase
correspondingly.} 
\label{spins}
\end{figure}
Schematically such a contact is shown on Fig. \ref{spins} in the
plane geometry. A transitional layer lies in the shaded area and
its structure may become complicated by a number of
reasons. First of all one should expect that the charge
effects discussed above for a contact between the two FM
phases exists at present case as well. 
Re-distribution of the carriers may shift the boundary between the magnetic phases.
Less clear is the magnetic structure of such a transitional layer: whether there is a sharp
boundary for the spins orientations on both sides of a contact or
the direction of spins may rotate going through the interface from
left to right, needs further discussion. In particular, since the energy of magnetic
anisotropy in manganites is rather weak (magnetization of ferromagnetic manganites is
saturated \cite{vonMolnar} at external fields of the order of 100 Oe) orientation
of moments in Fig. \ref{spins} may even change gradually forming a
structure similar to the one in the Bloch or Neel domain wall. The transitional region
between the two phases may include a canted A-phase,
magneto-conductivity of which has been studied above. 
It is highly desirable to further address these issues experimentally.
\section{Discussion and Summary}
We return to the applicability of our band model for different
doped manganites. Preparation of high-quality samples needs
significant efforts and therefore experiments with doped
manganites having good metallic conductivity at low temperatures
are rare in the literature. Ferromagnetic metallic LSMO samples
~\cite{Izumi,Quijada} with $\rho_0\simeq{10}^{-5}~\Omega\cdot{cm}$
have been reported. Estimates of Ref. [\onlinecite{GK}] give mean
free path values as large as $l\simeq{80a}$. More typical values
of residual resistivity are of the order of $1~m\Omega\cdot{cm}$
\cite{GK}, which means that the mean free path lies on the scale
of several inter-atomic distances. Higher values of resistivity
often correlate with larger mismatch in the ionic radii of the
constituent elements (tolerance factor $t<1$).

From various experimental data on A-phase in manganites
Pr$_{0.5}$Sr$_{0.5}$MnO$_3$, Nd$_{0.45}$Sr$_{0.55}$MnO$_3$ and
La$_{0.45}$Sr$_{0.55}$MnO$_3$ ~\cite{Izumi,Tokura2} the values of
the in-plane resistivity of these compounds lie in the range
$\rho\sim{3}\div{7}~(m\Omega{\cdot}cm)$. In these samples one
obviously has the conductivity regime in the A-phase lying close
to the mobility edge. As we have seen, magnetoconductivity and
other properties of the A-phase are interesting enough to justify 
the efforts to improve the values of resistivity in the A-phase
manganites. As for the LSMO films and hetero-structures prepared
in \cite{Izumi}, the value of residual resistivity in the
ferromagnetic state  of La$_{0.6}$Sr$_{0.4}$MnO$_3$ compound and
the value of the in-plane residual resistivity of the A-phase
La$_{0.45}$Sr$_{0.55}$MnO$_3$ compound are
$\rho_0=8\cdot{10^{-4}}~\Omega\cdot{cm}$ and
$\rho_0=3\cdot{10^{-3}}~\Omega\cdot{cm}$ respectively. The
films were prepared by a pulsed deposition method and obviously
are not clean enough. Nevertheless, qualitatively there is an agreement: 
our calculations of the values of
resistivity for ferromagnetic phase and A-phase (when $M=0$) for
the Sr concentrations $x=0.4$ and $x=0.5$ correspondingly gave a
factor of ten difference while the experimentally observed difference
is only a factor of two.

As far as an interface is concerned, the complications that lie on the 
theory part are as follows. The DE mechanism that played such an 
important role in exploring the
properties of the half-metallic magnetic ground state, bears a
non-local character. Therefore it is not straightforward to
account for it in an inhomogeneous problem with  a spatial
dependence near the interface (recently, there were 
attempts to present the DE mechanism in a local form to describe the various domain
structures in FM metallic manganites \cite{Golosov}). Secondly,
diminishing the number of band electrons, the anti-ferromagnetic 
super-exchange interaction between the $t_{2g}$ spins becomes comparable with the DE
interaction. The task of combining these two mechanisms to study an inhomogeneous
problem is already a serious problem. 

The A-phase ground state has been found in a number of other
compounds, such as Pr$_{0.5}$Sr$_{0.5}$MnO$_3$ and
Nd$_{0.45}$Sr$_{0.55}$MnO$_3$ \cite{Pr, Nd}. The transition into the A-phase
along the temperature axis is of the first order as it is expected from symmetry
considerations. It is accompanied by a change in the $c/a$
ratio \cite{Izumi,Tokura}. Such a transition is often described in the 
literature in terms of the "orbital ordering" \cite{Kugel}. 
We suggest to interpret this transition as a
cooperative Jahn-Teller effect involving the proper lattice distortion. 
Indeed, so far we have discussed changes
in the electronic band structure caused by spin re-arrangements only.
Meanwhile, as it was discussed in Introduction, lattice effects may also play an important
role. Judging from various experimental results (for a review, see \cite{Tokura2}),
the importance of the lattice effects may vary and depend on a specific compound.
For example, the lattice deformations strongly prevail in
Nd$_{0.45}$Sr$_{0.55}$MnO$_3$ compound \cite{Nd}. Its ground state shows huge anisotropy in
resistivity ($\rho_c/\rho_{ab}\sim 10^4$) and much lower in-plane conductivity compared to other
members of the A-phase family \cite{Izumi,Pr}. As it was shown in Ref. [\onlinecite{GK}]
the strong enough shear deformation $d_{x^2-y^2}$ of the oxygen octahedra ($c/a<1$) 
alone may lead to a in practically two-dimensional electronic spectrum. 
We suppose that Nd$_{0.45}$Sr$_{0.55}$MnO$_3$ is such 
an extreme case ($t\simeq{0.95}$) \cite{Tokura2}. 
After carriers are added to the 2D bands both their and core spins adjust themselves via DE and super-exchange
mechanisms. If in other A-phase compounds the tetragonal deformations of the lattice, $c/a<1$, is less 
pronounced, one may neglect it as we have done above for La$_{0.45}$Sr$_{0.55}$MnO$_3$.  
The phase diagram for the La$_{1-x}$Sr$_x$MnO$_3$ as a function of Sr doping $x$ and 
fixed tetragonal distortion $c/a$ has been theoretically studied in \cite{Fang}. We note, that 
the Jahn-Teller interaction has not been considered in \cite{Fang} and the electronic band structure 
was obtained for a fixed values of $c/a$.

Much attention in Ref. [\onlinecite{Izumi}] was also given to the characterization of 
properties of the hetero-structures consisting of mixed phases [F$_n$, A$_m$] 
($n$ and $m$ are the number of unit cells per period of the structure). 
Major conclusion drawn by the authors \cite{Izumi} from data on magnetization, 
structural characterization of the modulated films and their in-plane conductivity
have led them to the notion of stable FM and A-phase single layers, 
which preserve their integrity and stability even in very thin intervening structures such as 
$n,m = 2\div{5}$. This result supports the point that stability of the A-phase layers is due to the coherent
octahedra shrinkage, $c/a<1$, in the planes. Such an idea leaves room for speculation regarding 
the possible spin arrangements in the heterostructures. 

It is interesting that the conductivity measurements for the samples with composition [F$_{10}$, A$_{10}$]
gave lower values of conductivity as if it were due to the FM layers alone. We ascribe this to the effect
of charge re-distribution shown in Fig. \ref{PotEl}(b): ferromagnetic layers having more carriers supply part of them
to the A-phase layers. Ten layers of each phase is already a good approximation for the picture of a single
interface, as it is seen in Fig. \ref{spins}. 
With $n$ and $m$ decreased, the role of Coulomb effects becomes weaker and each layer preserves its
nominal composition and, hence, the in-plane conductivity.

One more comment we would like to add to the latter point concerns a pronounced increase in magneto-conductivity 
toward the [F$_{3}$, A$_{3}$] samples. If the number of electrons on each FM- and A-phase planes does not change,
the positive magneto-conductivity effect is an indication of stronger canting of the moments in these samples.
(Recall that large values of in-plane magneto-conductivity on Fig. \ref{cond}(a) is a result of changing Fermi
level at the transition between the 2D and 3D regimes).
Data provided by \cite{Izumi} qualitatively agree with our results for the values of in-plane conductivity for the
canted A-phase (Fig. \ref{cond}(a)).

To summarize, we have determined the electronic spectrum for the canted A-phase and calculated both conductivity
and magneto-conductivity of such a ground state. Calculations have been done in the framework of the two-band
model \cite{GK}. Disorder introduced by the substitutional doping results in an intimate correlation between
conductivity and number of carriers. Defects are described in terms of random Jahn-Teller centers.
Magneto-resistivity for both in-plane and out-of-plane directions are expressed in terms of magnetization, the latter
can be measured independently. Negative in-plane magneto-resistivity turns out to be large. 
Our study of effects at the contact between the
two phases showed rather large screening length of the order of five unit cells. We suggest that in the
transition to the A-phase the Jahn-Teller cooperative ordering plays a leading role. The resulting magnetic
structure determines the final electronic band spectrum. We applied the results
of the present paper to discuss the data obtained in \cite{Izumi}. There is a qualitative agreement 
of experimental data with our results and conclusions above. More experimental work is needed to further 
investigate interesting properties of the A-phase and heterostructures in the presence of an external 
magnetic field.
\section{Acknowledgments}
This work (M.D. and L.P.G.) was supported by the NHMFL through the
NSF cooperative agreement DMR-9527035 and the State of Florida
and by DARPA through the Naval Research Laboratory
Grant No. N00173-00-1-6005. The work of (V.Z.K.) was supported by 
DARPA under Contract No. 01J543.

\begin{appendix}
\section{Solution of the Poisson equation for the interface in the continuous model.}
Here we will present the exact solution of the Eq. (\ref{h-1}) in
the continuous limit. In that case, we can solve the Poisson equation
on each side of the contact will and, after using the boundary
conditions are able to obtain a general solution.

If the plane of the interface
coincides with $yz$ plane, the potential will depend on $x$
only. We assume the spectrum of the electrons to have a parabolic form:
$\varepsilon({k})={k^2/2m}$. The concentration of
electrons is given by:
\begin{equation}
n_{el}(x) =
\frac{(2m)^{3/2}}{3\pi^2\hbar^3}\left[\mu(x)\right]^{3/2},
\label{a-1s}
\end{equation}
where $\mu(x)$ is a local chemical potential. Taking into account
(\ref{a-1s}), we have to solve the Poisson equation:
\begin{equation}
\frac{d^2\Phi}{dx^2} = \frac{4{\pi}e^2}{t}\left[N_{Sr}(x) -
n_{el}(x)\right], \label{a-3}
\end{equation}
where $N_{Sr}$ is a Sr concentrations
defined by:
\begin{equation}
N_{Sr}(x) = \left\{
\begin{array} {r@{\quad}}
N_{L}, ~x\le{0}, \\
N_{R}, ~x\ge{0},
\end{array}
\right.
\end{equation}
We introduce the following notations: $\mu(x) + \Phi(x) = \zeta = const$, where $\zeta$ is an 
elechtro-chemical potential and $\Phi(x)$ is:
\begin{equation}
\Phi(x) = \left\{
\begin{array} {r@{\quad}}
\Phi_L(x), ~x\le{0}, \\
\Phi_R(x), ~x\ge{0},
\end{array}
\right.
\label{a-4}
\end{equation}

The boundary conditions for the potential preserving the charge conservation are:
\begin{equation}
\Phi_L(0) = \Phi_R(0), ~\frac{d\Phi_L}{dx} =~\frac{d\Phi_R}{dx}|_{x=0}.
\label{a-5}
\end{equation}
Introducing $V(x) = \zeta - \Phi(x) > 0$
and using (\ref{a-1s}) and (\ref{a-5}), Eq. (\ref{a-3}) is re-written as:
\begin{equation}
\frac{d^2V(x)}{dx^2} =
\frac{4\pi{e^2}(2m)^{3/2}}{3\pi^2\hbar^3}\cdot
\left\{V^{3/2}(x)-V^{3/2}(\pm\infty)\right\}, \label{a-6}
\end{equation}
where
\begin{equation}
\frac{4\pi{e^2}(2m)^{3/2}}{3\pi^2\hbar^3}\cdot{V^{3/2}(\pm\infty)} = N_{\pm}.
\nonumber
\end{equation}
The first integral of Eq. (\ref{a-6}) has the following form:
\begin{equation}
\begin{split}
\left(\frac{dv}{dx}\right) = \frac{1}{\kappa^2}
\left[\frac{2}{5}v^{5/2}(x) - v(x) + \frac{3}{5}\right], \\
v(x) = V(x)/V(\pm\infty),
\end{split}
\label{a-6p}
\end{equation}
where ${\kappa}^2\equiv{(3\pi\hbar^3)/(8e^2(2m)^{3/2})}$.
Equation (\ref{a-6p}) can be presented as:
\begin{equation}
\frac{dv}{\sqrt{(2/5)v^{5/2} - v + (3/5)}} = \pm\sqrt{2}{\kappa{dx}},
\label{a-7}
\end{equation}
where plus(minus) sign corresponds to $v_{L(R)}(x)$.
Integral on the left hand side of Eq. (\ref{a-7}) can be
calculated exactly. The final result reads:
\begin{eqnarray}
\Phi\left[V(x)\right] &=& \pm\frac{\kappa x}{\sqrt{10}} + C, \nonumber\\
\Phi\left[V(x)\right] &=& \sqrt{a_1 -
a_2x}\cdot\left\{2\sqrt{{a^*}_1 -
{a^*}_2x}\cdot(x+\alpha)\cdot{F}[\phi_1,m_1] - \right. \nonumber\\
&&\left. b_1\cdot\sqrt{a_2\cdot(x+\alpha)^2\cdot(b_2 +
b_3\cdot{x})}\cdot
\Pi[n_1;\phi_2,m_2]\right\}/\left\{\sqrt{(x+\alpha)\cdot(x^2+|\beta|^2)}\right\},
\label{a-9}
\end{eqnarray}
where F and $\Pi$ are elliptic integrals of the first and third
kind correspondingly and $\{a_i, b_i, \phi_i, m_i\}$ are complex
numbers. Making use of boundary conditions (\ref{a-5}) recovers
the results obtained numerically on Fig.\ref{PotEl}(a,b).

It is also useful to obtain the solution of the Poisson equation
(\ref{a-3}) in the linear approximation, assuming $\Phi(x)$ is
small compared to $\zeta$. After very simple algebra, (\ref{a-3})
takes the following form:
\begin{equation}
\frac{d^2\Phi(x)}{dx^2} = \frac{1}{\lambda^2}\left[\Phi(x) -
\Phi(\pm)\right], \label{a-10}
\end{equation}
where
\begin{equation}
\begin{split}
\frac{1}{\lambda^2} =
\left(\frac{2e^2(2m)^{3/2}}{\pi\hbar^3}\right)\cdot\zeta^{1/2},\\
\Phi(\pm\infty) = \frac{(2m)^{3/2}/(3\pi^2\hbar^3)\cdot\zeta^{3/2}
- N_{L(R)}}{2(2m)^{3/2}/(2\pi^2\hbar^3)\cdot\zeta^{1/2}}.
\label{a-11}
\end{split}
\end{equation}
Taking into account the boundary conditions (\ref{a-5}), the
integration of (\ref{a-11}) is straightforward. Thus the solution
of (\ref{a-11}) reads:
\begin{equation}
\Phi(x) = \left\{
\begin{array} {r@{\quad}}
\Phi(+\infty) + \delta\Phi\cdot{e^{-x/\lambda}}, ~x\ge{0}, \\
\Phi(-\infty) - \delta\Phi\cdot{e^{x/\lambda}}, ~x<{0},
\end{array}
\right. \label{afinal}
\end{equation}
where the notation $\delta\Phi =
0.5\cdot(\Phi(-\infty)-\Phi(+\infty))$ was introduced for brevity. As it is
easy to see, the result given by (\ref{afinal}) reproduces all the features of our
numerical solution shown in Fig. \ref{PotEl}(a,b).
\end{appendix}

\end{document}